\documentclass[runningheads]{llncs}
\usepackage{graphicx}
\begin{document}
\title{Automatically Identifying Political Ads on Facebook: Towards Understanding of Manipulation via User Targeting}
\titlerunning{Automatically Identifying Political Ads on Facebook}
%
\author{Or Levi\inst{1} \and
Sardar Hamidian\inst{2} \and
Pedram Hosseini\inst{2}}
\authorrunning{O. Levi et al.}
%
\institute{AdVerif.ai, Netherlands \email{or@adverifai.com} \and
The George Washington University, USA
\email{\{sardar,phosseini\}@gwu.edu}\\
 }
\maketitle              
\begin{abstract}
The reports of Russian interference in the 2016 United States elections brought into the center of public attention concerns related to the ability of foreign actors to increase social discord and take advantage of personal user data for political purposes. It has raised questions regarding the ways and the extent to which data can be used to create psychographical profiles to determine what kind of advertisement would be most effective to persuade a particular person in a particular location for some political event; Questions which have not been explored yet due to the lack of publicly available data. In this work, we study the political ads dataset collected by ProPublica, an American nonprofit newsroom, using a network of volunteers in the period before the 2018 US midterm elections. With the help of the volunteers, it has been made possible to collect not only the content of the ads but also the attributes that were used by advertisers to target the users. We first describe the main characteristics of the data and explore the user attributes including age, region, activity, and more, with a series of interactive illustrations. Furthermore, an important first step towards understating of political manipulation via user targeting is to identify politically related ads, yet manually checking ads is not feasible due to the scale of social media advertising. Consequently, we address the challenge of automatically classifying between political and non-political ads, demonstrating a significant improvement compared to the current text-based classifier used by ProPublica, and study whether the user targeting attributes are beneficial for this task. Our evaluation sheds light on questions, such as how user attributes are being used for political ads targeting and which users are more prone to be targeted with political ads. Overall, our contribution of data exploration, political ad classification and initial analysis of the targeting attributes, is designed to support future work with the ProPublica dataset, and specifically with regard to the understanding of political manipulation via user targeting.

\end{abstract}
\section{Introduction}

Social media platforms are collecting a great amount of personal user data. 
While the data can be used to improve the effectiveness of ad recommendation, as demonstrated by previous works \cite{Joshi_user,pandey_learning,grbovic_generating}, it also raises concerns related to user privacy, especially when it comes to political ads; Concerns, which have been amplified by the reports of Russian interference in the 2016 United States elections, when fake accounts linked to a Russian troll farm bought advertisements targeting millions of Facebook users prior to the election.
These concerns were further amplified by the Facebook-Cambridge Analytica data scandal, when it was revealed that Cambridge Analytica harvested the personal data of millions of people's Facebook profiles without their consent and used it for political purposes. 

Despite the growing public interest, the effect of political ad targeting on social media has not been explored yet due to the lack of publicly available data. Facebook has made available\footnote{\url{https://www.facebook.com/ads/archive/}} an archive of ads related to politics but that has included only the content of the ads.
In an effort to promote ad transparency and hold advertisers including political groups accountable, ProPublica, an American nonprofit newsroom, has collected a dataset of political ads in the period before the 2018 US midterm elections. Readers were asked to install a browser extension that automatically collected advertisements shown to them on Facebook without collecting personal information. With the help of the volunteers it has been made possible to collect not only the content of the political ads but also the attributes that were used by advertisers to target the users.
  
This work is the first to study the ProPublica political ads data and the use of targeting attributes, such as age, region, activity, and interests, for political advertising on social media. First, we describe the main properties of the dataset and provide a series of interactive illustrations by leveraging the targeting attributes, in addition to election information collected from online resources.

Second, in order to study the potential to manipulate users for political purposes via the targeting attributes, it is important to initially identify which ads, and advertisers, are politically oriented. We are motivated by the increasing efforts of both social media platforms, and investigative journalism organizations, to improve the transparency and scrutiny around political advertising and study their effect on the spread of misinformation and social discord. However, given the large scale of social media advertising, manually checking ads is impractical.

Consequently, we address the challenge of automatically classifying between political and non-political ads.
While the data released by ProPublica contains only ads that were identified as political by an existing classifier, we notice there is still a great amount of disagreement compared to the judgments by the volunteers, and aim to improve the text classification.

In addition to identifying language differences, we also consider the following research question: can the targeting attributes be used for identification of political ads? In other words, are there differences in the patterns of user targeting between political and non-political ads? The evaluation of our method sheds light on how user attributes are being used for political ads targeting and what kind of user profiles are more likely to be targeted.
For instance, we find that political advertisers are more likely to use location targeting, and that users in battleground states are more likely to be targeted with political ads.

\section{Related Work}

Previous works have demonstrated the effectiveness of targeting attributes for ad recommendation, based on user behavior \cite{pandey_learning,grbovic_generating}, user demographics \cite{jansen_gender} or a combination of the two. For instance, Bagherjeiran et al. \cite{Joshi_user} proposed to build a generic  user  profile  with  demographic  and  behavioral  information about  the  user, and learned a mapping from non-textural user features to the textual space of ads that helped to improve the click rate on ads.


Another related line of work is the classification of political orientation from text on social media.
\cite{hoang_politics,volkova_inferring,maynard_automatic}. Pennacchiotti et al. \cite{pennacchiotti_democrats} proposed to automatically construct user profiles, to identify the political affiliation of users, based on features related to profile information, messaging behavior, linguistic content and social connections. Similarly, Boutet et al. \cite{boutet_what} used the number of Twitter messages referring to a particular political party to identify the political leaning of users. In this work we focus on a different task. Rather than identifying a political orientation, we aim to distinguish between political and non-political ads. For this task, we use the novel targeting attributes, that were used by advertisers to target users and have been made available only recently, with the release of the ProPublica dataset.

\section{The ProPublica Dataset}

\begin{figure}[]
\centering
   \includegraphics[scale=0.35]{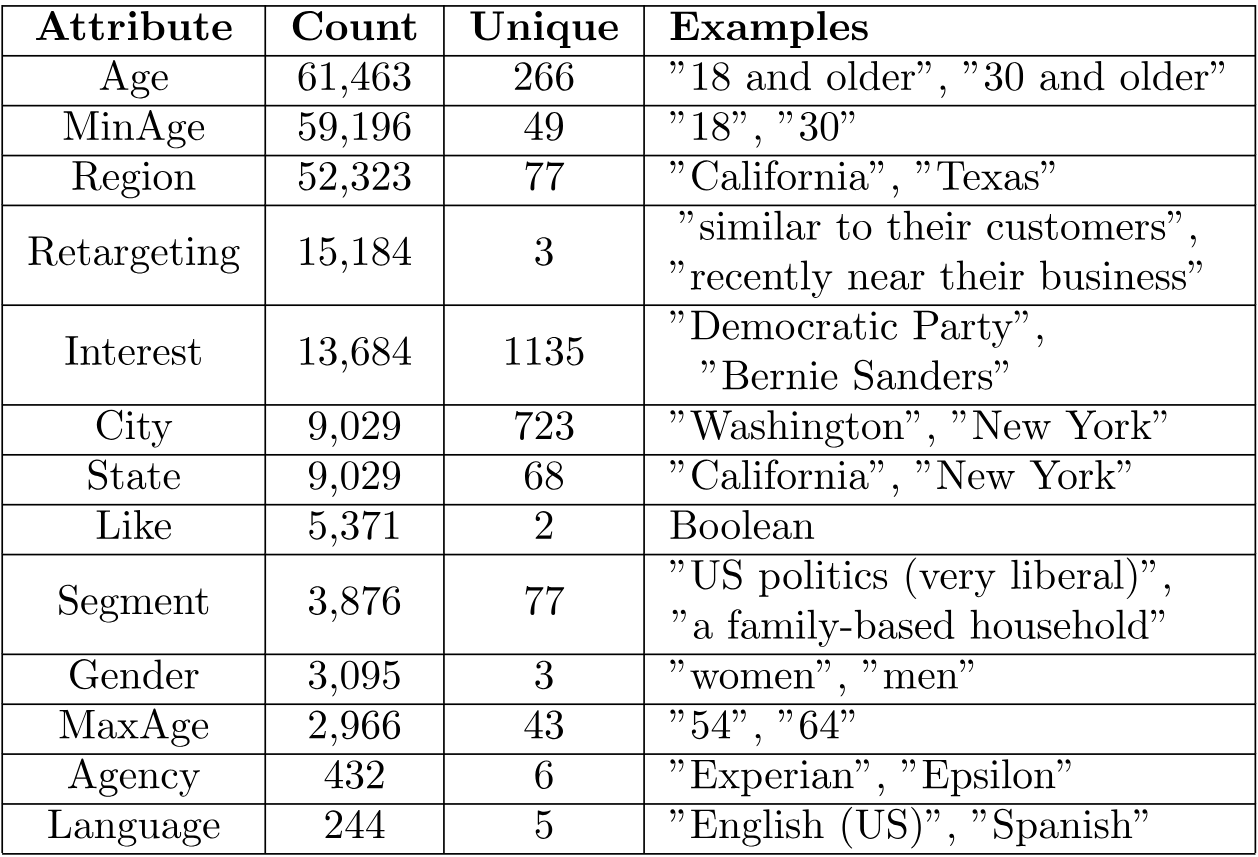}
   \caption{The Targeting Attributes.  For each attribute we present the number of occurrences in the data, the number of unique values and a couple of examples.}
    \label{fig:table1}
\end{figure}

The ProPublica political ads dataset\footnote{\url{https://propublica.org/datastore/dataset/political-advertisements-from-facebook}} includes information regarding the content of the ads, such as title, message and images; the number of users who voted it as political or not political; and the targeting attributes, as described in figure \ref{fig:table1}. Overall, the data includes more than 68,000 ads from 5,700 different advertisers collected in the period between August 2017 and October 2018.

\begin{figure}[]
 \centering
   \includegraphics[scale=0.95]{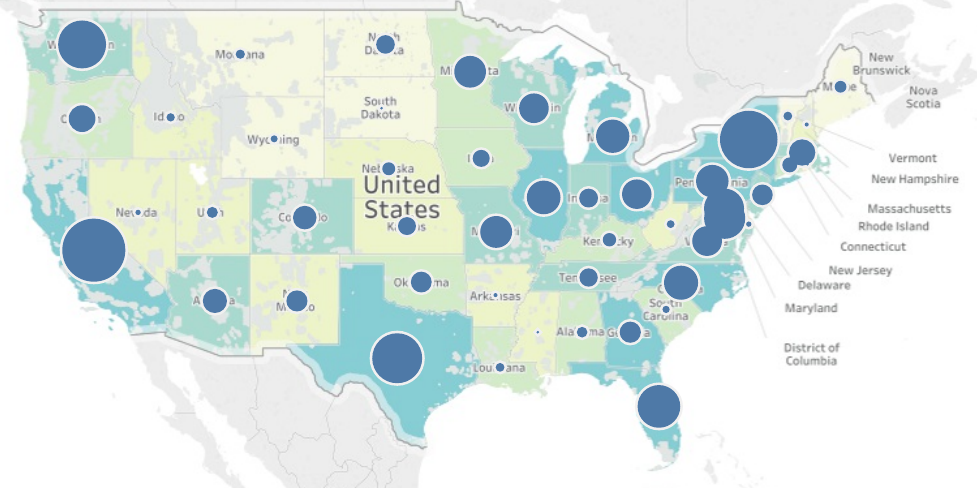}
   \caption{The distribution of political ads on Facebook in different states based on population. Circle size is the percentage calculated by number of ads in the state divided by total number of ads in the US.}
    \label{fig:top10interest}
\end{figure}

To manifest a better insight into the properties of the data, we provide a series of interactive illustrations\footnote{\url{https://tabsoft.co/2RErMBD}} by leveraging the targeting attributes, in addition to election information collected from online resources.
Figure \ref{fig:top10interest} is one of the graphs from this dashboard illustrating the distribution of political ads on Facebook based on geographical information collected from the targeting attributes. According to the map, users in highly populated states (Darker green relative to high regional population) like California, New York, Texas, and Washington are more prone to be targeted by political ads. 

\begin{figure}[]
   \centering
   \includegraphics[scale=0.3]{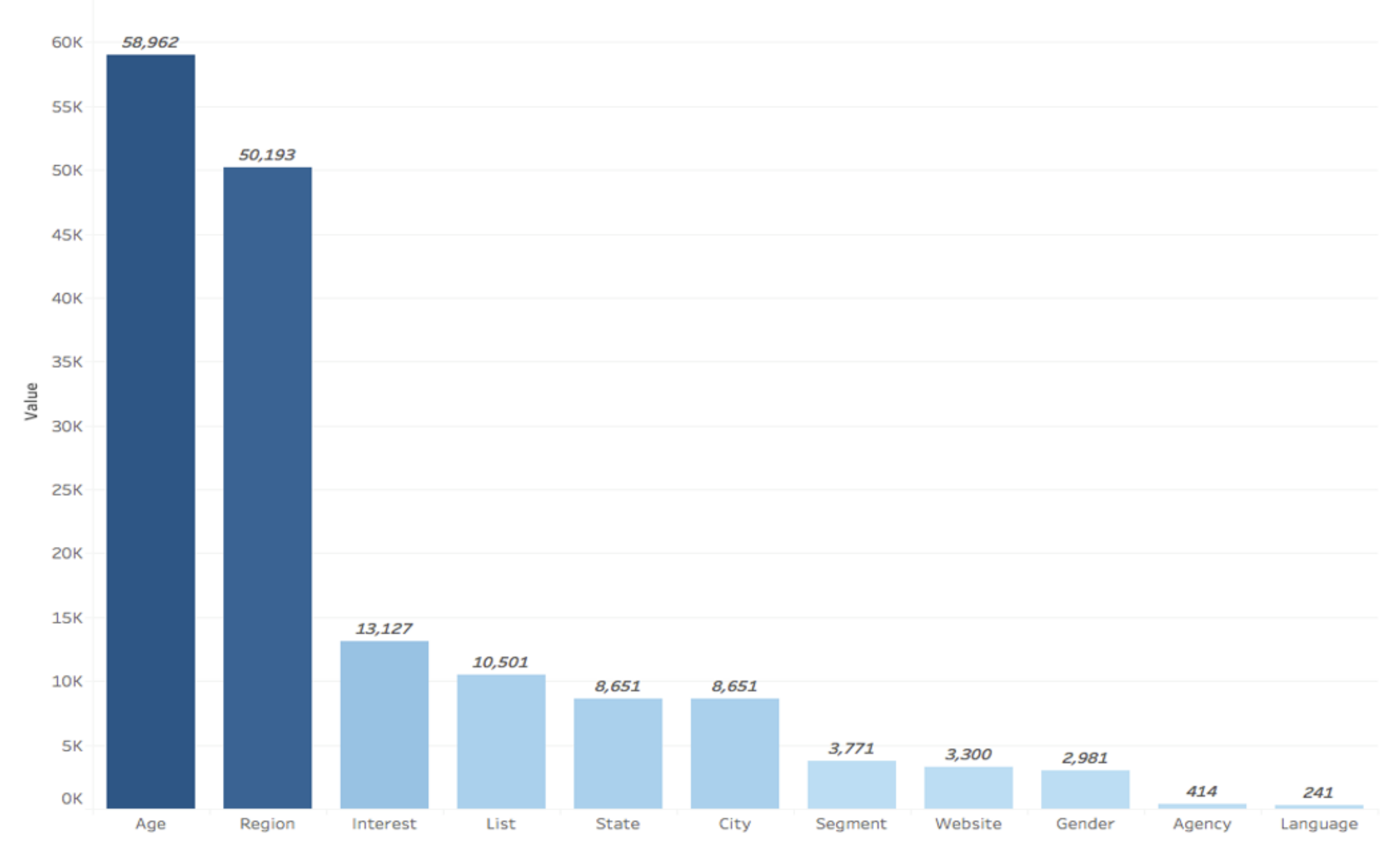}
   \caption{Number of political ads for each of the targeting attributes.}
    \label{fig:politicalfortargeting}
\end{figure}

Figure \ref{fig:politicalfortargeting} shows the number of political ads for each of the targeting attributes. According to this chart, the top two targeting attributes used in political ads are the age and the region of the Facebook users. More than 70 percent of the time Facebook users are targeted by political ads is because they meet a certain age and location criteria, as opposed to language, agency, and gender with only 2 percent.

\begin{figure}[]
 \centering
   \includegraphics[scale=0.6]{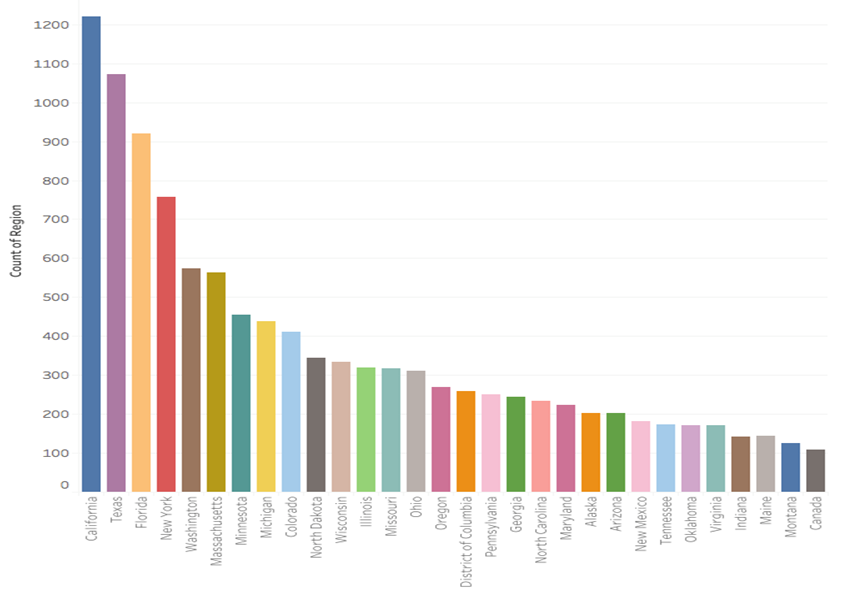}
   \caption{The distribution of political ads in different states.}
    \label{fig:states}
\end{figure}

Region is the second most important attribute used in political ads. As shown in figure \ref{fig:states}, Facebook users in California, Texas, Florida and New York are almost 10 times more likely to be targeted with political ads than states like Indiana, Montana or even Virginia.

\begin{figure}[]
 \centering
   \includegraphics[scale=0.4]{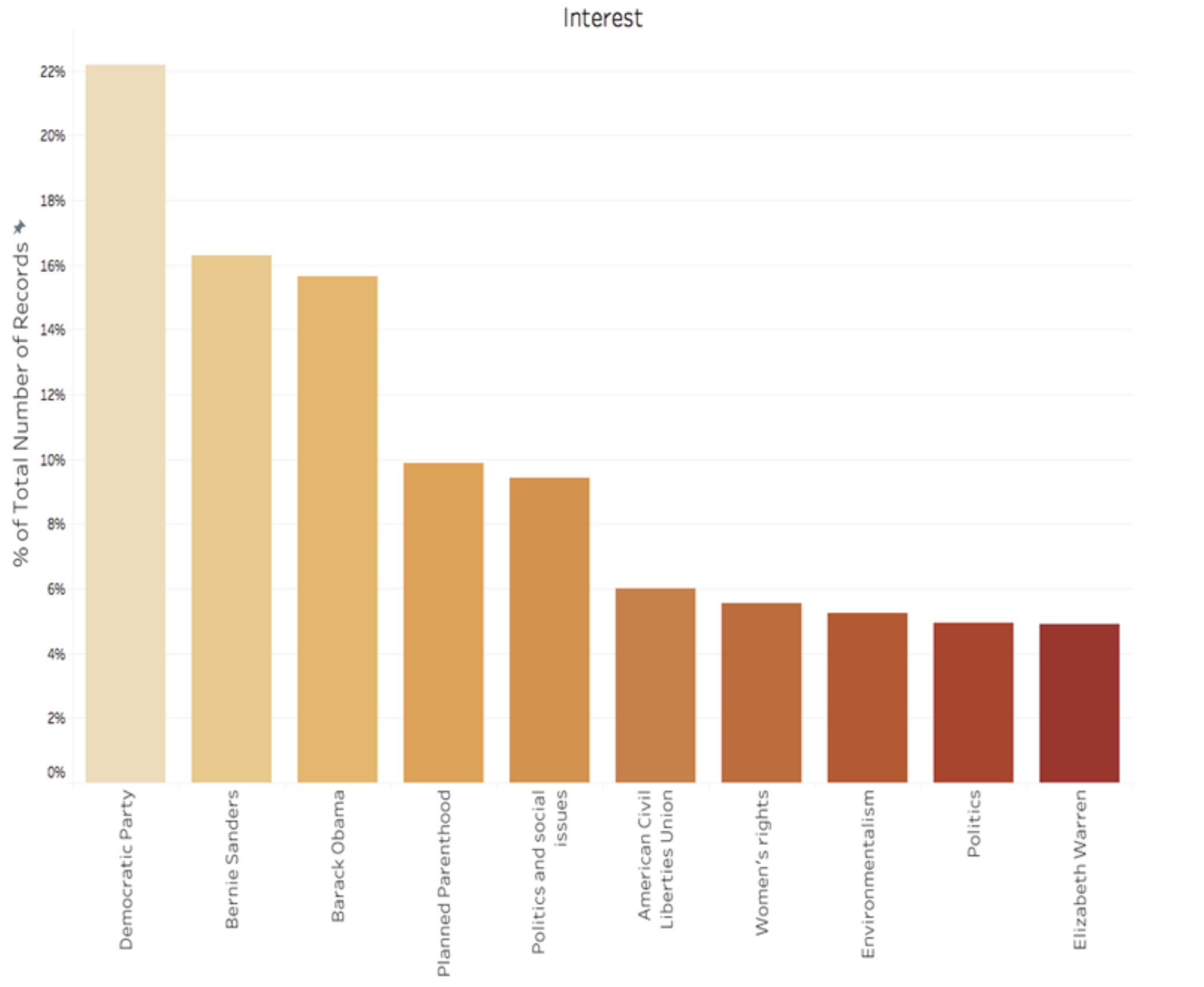}
   \caption{Top 10 Interest topics used for targeting the political ads.}
    \label{fig:interest}
\end{figure}

After age and region, interest is the third most important attribute for political ad targeting. Figure \ref{fig:interest} shows the top 10 interest topics used by advertisers. According to the chart, Facebook users with interest in the "Democratic Party", "Bernie Sanders" and "Barack Obama" are more prone to be targeted by political ads than the other interest topics.



\begin{figure}[]
 \centering
   \includegraphics[scale=0.5]{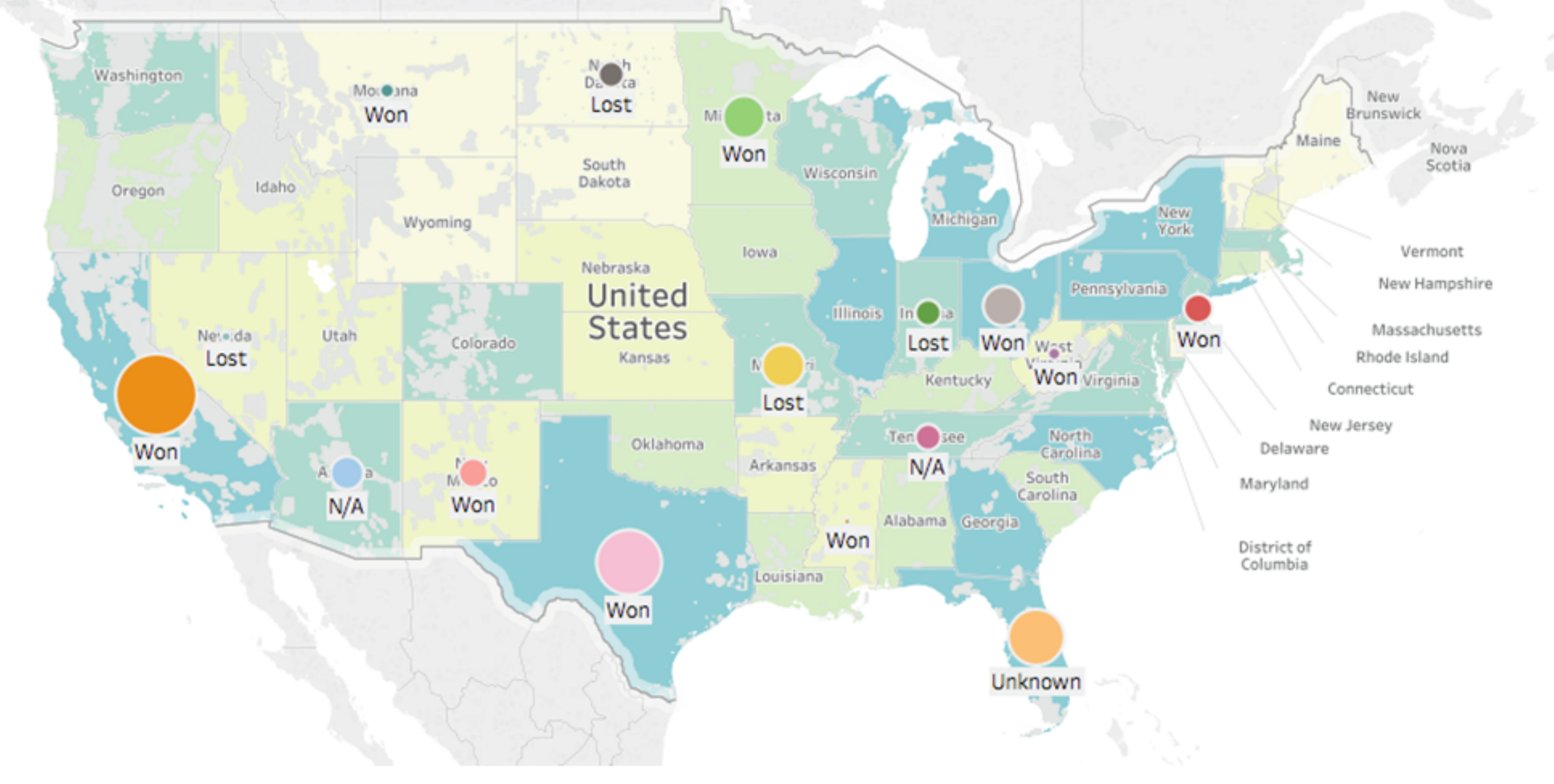}
   \caption{Incumbent status for battleground states vs. the volume of political advertising, represented by the circle size.}
    \label{fig:incumbent}
\end{figure}

Figure \ref{fig:incumbent} shows the number of political ads for each of the battleground states, in addition to the election outcome on the map.  There appears to be no significant correlation between the election outcome and the number of ads in battleground states.

\section{Method}
To study the effects of political manipulation via user targeting, we first address the challenge of automatically classifying between political and non-political ads. The classification labels are based on the 'political' and 'not political' fields in the data, which reflect the number of volunteers who have voted an ad as political or not political. Ads with more 'political' votes are classified as political and vice versa. We disregard ads with equal amounts of 'political' and 'not political' votes.

To classify political ads, ProPublica have been using a text classifier, such that the dataset contains only ads that were identified as political with a probability greater than 70\% (see 'political probability' field). However, a quick examination of the probabilities assigned by the classifier compared to the judgments by the volunteers shows still a great amount of disagreement. For instance, there are examples where the classifier picks up on a keyword like 'vote' but it is used in a non-political context. We hypothesize that using bigrams together with a tree-based classifier could help with these false positives and improve the performance of the classifier. A key consideration is also to provide a simple method that will be computationally inexpensive. 

Given that the data made available by ProPublica contains only ads that were already identified by the current classifier, political ads, as judged by the volunteers, outnumber non-political ads with a 9:1 ratio. To address this challenge we use an imbalance correction method, giving a penalty to the over-represented class, with a weight that is inversely proportional to the class frequencies in the input data:
\begin{equation}
Weight(y)= \frac{n\_samples}{n\_samples(y)}
\end{equation}
where
\begin{math}
n\_samples
\end{math}
and 
\begin{math}
n\_samples(y)
\end{math}
is the number of samples in general, and from class
\begin{math}
y
\end{math}
, respectively. 

We next turn to study our research question with regard to the potential of the targeting attributes to help with identifying political ads. The 'targets' field holds the targeting attributes of each ad. As part of the data pre-processing, we transform this field into separate columns, each representing one of the targeting attributes. Since the 'Region' and 'State' attributes are mostly overlapping, we drop the 'State' and use the 'Region', which occurs in more entries. We drop the sparse attributes 'Engaged with Content', with only 9 entries, and 'Language', with only 4 Non-English entries. Instead of the 'Age' attribute, which represents the targeted age range, we use the 'MinAge' and 'MaxAge' attributes, which represent the range limits. All the attributes are treated as categorical variables and transformed using one-hot encoding, except for the numerical attributes 'MinAge' and 'MaxAge'. Note that this still supports cases of users with multiple values for the same attribute, e.g. multiple interests, given that each interest is represented by a separate binary feature.

The ad text is obtained by concatenating the 'title' and 'message' fields of the ad. We use a TF-IDF vector representation as implemented by the sci-kit learn toolkit with Snowball stemming and stop words removed.

The baseline method by ProPublica uses a Multinomial Naive Bayes classifier. For the tree-based classification model, we use the Gradient Boosting Decision Tree (GBDT) as implemented by the LightGBM toolkit \cite{ke_LightGBM}. We test two methods, with the text only and together with the targeting attributes. The model hyper-parameters are tuned using a five-fold grid search cross validation.

\section{Evaluation}

We evaluate the performance of our method for political ad classification using the F1 measure.
We split the data into a train and held-out test sets. To prevent over-fitting on patterns of specific advertisers, we separate the data such that each advertiser is either in the train or the test set. We randomly sample 20\% of the advertisers and the ads of these advertisers are used for the test set only.

Table \ref{table:table1} shows the main results. Our method, that employs bigrams and a Gradient Boosting Machine classifier, outperforms the Multinomial Naive Bayes classifier currently used by ProPublica, with a significant increase in the F1 measure. To test for statistical significance, we use the paired bootstrap test as recommended by Reichart et al. \cite{dror_recommended}. With the bootstrap test, we draw 1000 different samples. The size of each sample is the same as the full data, and the train and test sets are obtained using the above-mentioned split by advertisers. For each sample we evaluate the F1 score of the baseline and our method. The scores are then used to check the statistical significance via the bootstrap test implemented by Dror et al.\footnote{\url{https://github.com/rtmdrr/}} with a 0.05 significance level.

\begin{table*}[t]
 \centering
\small
\caption{Main Results. Our method, with the Gradient Boosting Machine classifier, achieves significant improvement on the F1 measure compared to the existing ProPublica classifier. Also, using the targeting attributes outperforms the text only based methods. Bold: best result among methods. Statistically significant differences with the ProPublica baseline and the GBM text only classifier are  marked with '*' and '**', respectively.}
\begin{tabular}{|c|c|c|c|}
\hline
\textbf{Method}             & \textbf{Precision} & \textbf{Recall} & \textbf{F1}     \\ \hline
MultinomialNB: Text Only (ProPublica)                   & 88.75              & 96.65           & 92.53           \\ \hline
GBM: Text Only                   & 90.33              & 99.25           & 94.58*           \\ \hline
GBM: Text + Targeting attributes & \textbf{90.83}              & \textbf{99.68}           & \textbf{95.05**} \\ \hline
\end{tabular}
\label{table:table1}
\end{table*}

Further to our research question, the evaluation also shows that using the targeting attributes for classification of political ads can further improve the performance, compared the text-only methods. Even though the improvement is not large, it gives motivation to further investigate differences in the patterns of targeting users between political and non-political ads.

\begin{figure}[]
 \centering
   \includegraphics[scale=0.4]{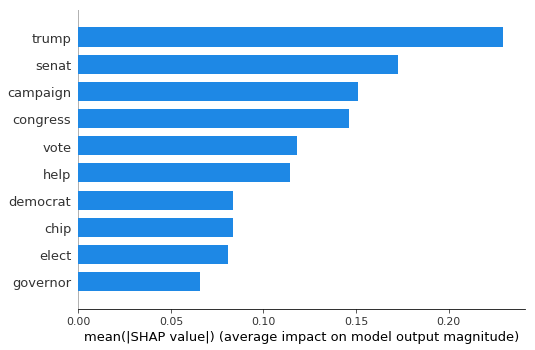}
   \caption{Top 10 Most Important Keywords. The list contains terms that can be expected to be associated with political ads.}
    \label{fig:top_keywords}
\end{figure}

To study the feature importance to our LightGBM model, we use Tree SHAP \cite{lundberg_consistent}, a fast algorithm to compute SHAP values \cite{lundberg_unified} for trees, as implemented by Lundberg et al.\footnote{\url{https://github.com/slundberg/shap}}. Figure \ref{fig:top_keywords} shows the most important keywords, sorted by the sum of SHAP value magnitudes over all training samples. The list includes terms that can be expected to be associated with political ads, such as "trump", "senate", "congress" and more.

\begin{figure}[]
 \centering
   \includegraphics[scale=1]{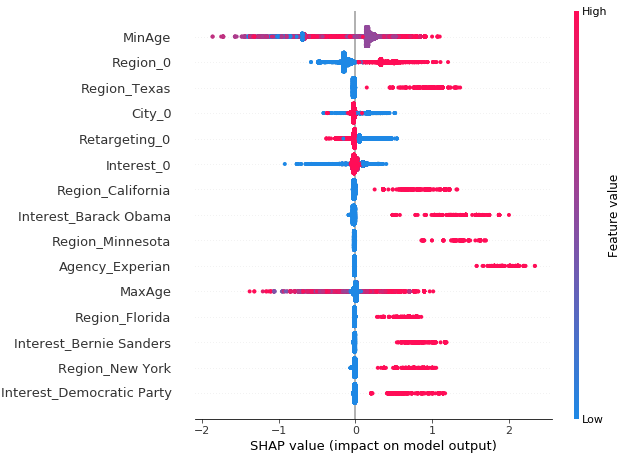}
   \caption{Top 15 Most Important Targeting Attributes. SHAP  values  show  the  distribution  of  the  impacts  each  feature  has  on  the model  output. The '\_0' notation is used for features representing an attribute with a missing value. We observe that certain Regions and Interests are more likely to be targeted with political ads.}
    \label{fig:top_attributes}
\end{figure}

Figure \ref{fig:top_attributes} shows the most important targeting attributes. It uses SHAP values to show the distribution of the impacts each feature has on the model output. The color represents the feature value: high (red) or low (blue), which is simply 1 or 0 for the binary attributes.

 The most important attribute is the 'MinAge'. We can see that above a certain threshold, higher age values increase the chance of seeing a political ad. Further examination (not presented herein) reveals that this threshold corresponds to '18', which is also the legal voting age in the US. We can also see that users with interest related to 'Barack Obama', 'Bernie Sanders' or the 'Democratic Party' are more likely to see political ads. Lastly, this analysis reveals that non-political advertisers are less likely to use the 'Region' attribute for targeting. This could be expected since politicians are more likely to target the state that elects them. On the other hand,
 users located in 'Texas', 'California', 'Minnesota', 'Florida' and 'New York' are more likely to be targeted with political ads. A comparison with the list provided by Ballotpedia.org \footnote{\url{https://ballotpedia.org}} reveals that all the states except 'New York' are considered as battleground states. 

\section{Conclusion and Future Work}

This work is the first to study the ProPublica political ads dataset. The uniqueness of the data lies in the targeting attributes that that were used by advertisers to target users on social media. We first described the main characteristics of the data and explored the targeting attributes with a series of interactive illustrations.
Then, as a first step towards understating of political manipulation via user targeting, we addressed the challenge of automatically identifying political ads. Our method outperformed the current text-based classifier used by ProPublica with a significant improvement in the F1 measure. We also demonstrated the potential for further improvement in identifying political ads by using the targeting attributes. Lastly, we studied the feature importance of our method and pointed out interesting insights with regard to language differences between political and non-political ads, and the use of targeting attributes in political advertising, such as that users in battleground states are more likely to be targeted.

We consider several avenues for future work. First, the dataset contains additional information that has not been utilized in this work. For example, the ad images could potentially be helpful for the classification of political ads. A key consideration for us has been to provide a simple and scalable solution. This leaves room for future work to experiment with more sophisticated methods, such as learning user-based embeddings based on the targeting attributes to potentially show even greater improvement in performance compared to the text-only methods. 
Moreover, the identification of political ads allows for future work to explore the rich data provided by the targeting attributes in more detail. For example, to investigate which political ads were associated with which users and which targeting attributes, and specifically with regard to regional targeting which we found to be important. Overall, we hope these preliminary results will help to spark future work on understanding of political manipulation via user targeting and ways of addressing it.

%
%
%
%

\end{document}